\documentclass[11pt,a4paper]{article}

\usepackage{fontspec}
\usepackage[titletoc,title,header]{appendix}

\usepackage{amsmath}
\usepackage{amssymb}
\usepackage{amsthm}
\usepackage{mathtools}
\usepackage{xparse}    

\usepackage[margin=1in]{geometry}

\usepackage{graphicx}
\usepackage{xcolor}

\usepackage{hyperref}
\hypersetup{
  colorlinks=true,
  linkcolor=blue,
  citecolor=blue,
  urlcolor=blue,
  pdfauthor={Yoshitsugu Sekine},
  pdftitle={Magnetism and infrared divergence in a Hubbard-phonon interacting system}
}

\usepackage[numbers]{natbib}
\theoremstyle{plain}
\newtheorem{thm}{Theorem}[section]
\AtEndEnvironment{thm}{\qed}

\AtEndEnvironment{lem}{\qed}
\newtheorem{prop}[thm]{Proposition}
\AtEndEnvironment{prop}{\qed}

\AtEndEnvironment{cor}{\qed}

\AtEndEnvironment{conj}{\qed}

\AtEndEnvironment{fact}{\qed}

\theoremstyle{definition}
\newtheorem{defn}[thm]{Definition}
\AtEndEnvironment{defn}{\qed}

\AtEndEnvironment{ex}{\qed}

\theoremstyle{remark}

\AtEndEnvironment{rem}{\qed}

\makeatletter
\providecommand*{\dashv}{\mathrel{\mathpalette\@Dashv\vDash}}
\newcommand*{\@dashv}[2]{\reflectbox{$\m@th#1#2$}}
\makeatother
\newcommand{\abs}[1]{\left| #1 \right|} 
\NewDocumentCommand{\weaknorm}{O{\dbk} m}{#1{#2}} 
\newcommand{\norm}[1]{\left\Vert #1 \right\Vert} 

\newcommand{\bkt}[2]{\left\langle #1,\,#2 \right\rangle} 
\newcommand{\rbkt}[2]{\left( #1,\,#2 \right)} 
\newcommand{\isomto}{\mathrel{\rightarrowtail\kern-1.9ex\twoheadrightarrow}} 
\newcommand{\cbk}[1]{\left\{ #1 \right\}} 
\newcommand{\dbk}[1]{\left\langle #1 \right\rangle} 
\newcommand{\rbk}[1]{\left( #1 \right)} 
\newcommand{\sqbk}[1]{\left[ #1 \right]} 
\newcommand{\fun}[2]{#1 \rbk{#2}} 

\NewDocumentCommand{\imunit}{O{\mathsf{i}}}{#1} 
\NewDocumentCommand{\placeholder}{O{\bullet}}{#1} 
\NewDocumentCommand{\trace}{O{\operatorname{Tr}}}{#1} 
\newcommand{\Ker}{\operatorname{Ker}} 
\newcommand{\dom}{\operatorname{dom}} 
\newcommand{\eqcsq}[1]{\sqbk{#1}} 
\newcommand{\idone}{1} 
\newcommand{\kroneckerdelta}{\delta} 
\newcommand{\laplacian}{\Delta} 
\newcommand{\napiernum}{\mathsf{e}} 
\newcommand{\opchern}[1]{\operatorname{ch}} 
\newcommand{\opimag}{\operatorname{Im}} 
\newcommand{\opreal}{\operatorname{Re}} 
\newcommand{\setSymbolDownLeft}[2]{{\vphantom{#2}}_{#1}{#2}} 
\newcommand{\setSymbolUpLeft}[2]{{\vphantom{#2}}^{#1}{#2}} 
\DeclareMathOperator*{\algotimes}{\hat{\otimes}} 
\NewDocumentCommand{\agvariety}{O{\mathcal}}{#1} 
\NewDocumentCommand{\cmdrel}{O{\omega}}{#1} 
\NewDocumentCommand{\dfsp}{O{A}}{#1} 
\NewDocumentCommand{\eqcpointed}{O{\eqcsq} m}{#1{#2}_{\ast}} 
\NewDocumentCommand{\fnheaviside}{O{H}}{#1} 
\NewDocumentCommand{\fthol}{O{\mathcal{O}}}{#1} 
\NewDocumentCommand{\ftmero}{O{\mathcal{M}}}{#1} 
\NewDocumentCommand{\grcentralizer}{O{Z}}{#1} 
\NewDocumentCommand{\grmetform}{O{2} m m}{\grmet[#1] \! \rbkt{#2}{#3}} 
\NewDocumentCommand{\grmet}{O{2}}{\setSymbolDownLeft{#1}{g}} 
\NewDocumentCommand{\grnormalizer}{O{N}}{#1} 
\NewDocumentCommand{\gropasym}{O{A}}{#1} 
\NewDocumentCommand{\gropsym}{O{S}}{#1} 
\NewDocumentCommand{\grpermorderedpair}{O{\mathcal{P}}}{#1} 
\NewDocumentCommand{\grsym}{O{\mathfrak{S}} m}{#1_{#2}} 
\NewDocumentCommand{\gtbase}{O{\mathcal}}{#1} 
\NewDocumentCommand{\gtfilter}{O{\mathcal}}{#1} 
\NewDocumentCommand{\gtfmlclosed}{O{\mathcal}}{#1} 
\NewDocumentCommand{\gtfmlopen}{O{\mathcal}}{#1} 
\NewDocumentCommand{\gtopenball}{O{U}}{#1} 
\NewDocumentCommand{\gtopencover}{O{\mathcal}}{#1} 
\NewDocumentCommand{\gtopennbh}{O{\mathcal}}{#1} 
\NewDocumentCommand{\gtpreopencover}{O{\mathcal}}{#1} 
\NewDocumentCommand{\gtsubbase}{O{\mathcal}}{#1} 
\NewDocumentCommand{\gtvicinity}{O{\mathcal}}{#1} 
\NewDocumentCommand{\lasp}{O{\mathcal}}{#1} 
\NewDocumentCommand{\latprightrbk}{O{\top} m}{\rbk{#2}^{#1}} 
\NewDocumentCommand{\latpright}{O{\top} m}{#2^{#1}} 
\NewDocumentCommand{\latp}{O{t} m}{\setSymbolUpLeft{#1}{#2}} 
\NewDocumentCommand{\lpdistribution}{O{\mu} m}{#2_{\ast,#1}} 
\NewDocumentCommand{\lpmollifier}{O{\rho}}{#1} 
\NewDocumentCommand{\lpofpositive}{O{\chi}}{#1} 
\NewDocumentCommand{\manliederiv}{O{L}}{#1} 
\NewDocumentCommand{\mansmoothnbh}{O{\mathcal}}{#1} 
\NewDocumentCommand{\mblfmldsysgenerated}{O{d} m}{\fun{#1}{#2}} 
\NewDocumentCommand{\mblfmlgenerated}{O{\sigma} m}{\fun{#1}{#2}} 
\NewDocumentCommand{\oacorrfn}{O{\Gamma}}{#1} 
\NewDocumentCommand{\oagnsvector}{O{\Omega}}{#1} 
\NewDocumentCommand{\oaideal}{O{\mathcal}}{#1} 
\NewDocumentCommand{\oanumberoperator}{O{A}}{#1} 
\NewDocumentCommand{\oaposcone}{O{\mathcal{P}}}{#1} 
\NewDocumentCommand{\oapressure}{O{P}}{#1} 
\NewDocumentCommand{\oarepn}{O{\pi}}{#1} 
\NewDocumentCommand{\oaspkmsstate}{O{\txtkms}}{#1} 
\NewDocumentCommand{\oaspnormalstate}{O{N}}{#1} 
\NewDocumentCommand{\oasppurestate}{O{P}}{#1} 
\NewDocumentCommand{\oaspstate}{O{E}}{#1} 
\NewDocumentCommand{\oastatevector}{O{\Omega}}{#1} 
\NewDocumentCommand{\oastate}{O{\omega}}{#1} 
\NewDocumentCommand{\opdilation}{O{\delta}}{#1} 
\NewDocumentCommand{\opdmat}{O{\rho}}{#1} 
\NewDocumentCommand{\opfockan}{O{a}}{#1} 
\NewDocumentCommand{\opfockcran}{O{a}}{#1^{\#}} 
\NewDocumentCommand{\opfockcrdagger}{O{a}}{#1^{\dagger}} 
\NewDocumentCommand{\opfockcr}{O{a}}{#1^{\ast}} 
\NewDocumentCommand{\opfocknumber}{O{N}}{#1} 
\NewDocumentCommand{\opfocksegalconj}{O{\pi}}{#1} 
\NewDocumentCommand{\opfocksegal}{O{\phi}}{#1} 
\NewDocumentCommand{\opfockvac}{O{\Omega}}{#1} 
\NewDocumentCommand{\opspecmeas}{O{E}}{#1} 
\NewDocumentCommand{\opspec}{O{} m}{\fun{\sigma_{#1}}{#2}} 
\NewDocumentCommand{\optransl}{O{\tau}}{#1} 
\NewDocumentCommand{\opvarspec}{O{} m}{\operatorname{spec}_{#1} #2} 
\NewDocumentCommand{\physaction}{O{\mathcal{A}}}{#1} 
\NewDocumentCommand{\physcharge}{O{e}}{\mathrm{#1}} 
\NewDocumentCommand{\physcplconst}{O{\mathsf{g}}}{#1} 
\NewDocumentCommand{\physelectrostaticcapasity}{O{\mathrm{Cap}}}{#1} 
\NewDocumentCommand{\physenergy}{O{E}}{#1} 
\NewDocumentCommand{\physgse}{O{E}}{#1_{0}} 
\NewDocumentCommand{\physham}{O{H}}{#1} 
\NewDocumentCommand{\physlagdensity}{O{\mathcal{L}}}{#1} 
\NewDocumentCommand{\physlag}{O{L}}{#1} 
\NewDocumentCommand{\physliouvilean}{O{L}}{#1} 
\NewDocumentCommand{\physmass}{O{m}}{#1} 
\NewDocumentCommand{\prbbrownmv}{O{B}}{#1} 
\NewDocumentCommand{\prbcharfun}{O{\chi}}{#1} 
\NewDocumentCommand{\prbdist}{O{\mathcal{P}}}{#1} 
\NewDocumentCommand{\prbgaussianmeasure}{O{\msrcal{N}}}{#1} 
\NewDocumentCommand{\prbnormaldist}{O{N}}{#1} 
\NewDocumentCommand{\prbpoissonprocess}{O{N}}{#1} 
\NewDocumentCommand{\prbprocess}{O{X}}{#1} 
\NewDocumentCommand{\prbqspace}{O{\mathcal{Q}}}{#1} 
\NewDocumentCommand{\prbspsample}{O{\Omega}}{#1} 
\NewDocumentCommand{\prbspstate}{O{\mathsf{S}}}{#1} 
\NewDocumentCommand{\psh}{O{\mathfrak}}{#1} 
\NewDocumentCommand{\qtquantumchannel}{O{\mathcal{L}}}{#1} 
\NewDocumentCommand{\repn}{O{\pi}}{#1} 
\NewDocumentCommand{\schattencls}{O{\mathbb{K}}}{#1} 
\NewDocumentCommand{\setfmlcylinder}{O{\mathcal{C}}}{#1} 
\NewDocumentCommand{\setfml}{O{\mathcal}}{#1} 
\NewDocumentCommand{\setindex}{O{\mathcal} m}{#1{#2}} 
\NewDocumentCommand{\setlattice}{O{\Gamma}}{#1} 
\NewDocumentCommand{\setspecial}{O{\mathcal} m}{#1{#2}} 
\NewDocumentCommand{\shdiffform}{O{\sheaf{A}}}{#1} 
\NewDocumentCommand{\sheaf}{O{\mathfrak}}{#1} 
\NewDocumentCommand{\smchemicalpotential}{O{\mu}}{#1} 
\NewDocumentCommand{\smenergydensity}{O{\varrho}}{#1} 
\NewDocumentCommand{\smfluctuationwithdmat}{O{\beta} m}{\smuncertaintywithdmat[#1]{#2}^2} 
\NewDocumentCommand{\sminvtemperature}{O{\beta}}{#1} 
\NewDocumentCommand{\smlocaldensityoperator}{O{\rho}}{#1} 
\NewDocumentCommand{\smmicrocanonicalstate}{O{\beta} m}{\physmean{#2}_{#1}} 
\NewDocumentCommand{\smnumberdensity}{O{\rho}}{#1} 
\NewDocumentCommand{\smooth}{O{\mathcal{E}}}{#1} 
\NewDocumentCommand{\smparticlenumber}{O{N}}{#1} 
\NewDocumentCommand{\smpressure}{O{p}}{#1} 
\NewDocumentCommand{\smspecificfreeenergy}{O{\bar{f}}}{#1} 
\NewDocumentCommand{\smthermalvac}{O{\beta}}{\Omega_{#1}} 
\NewDocumentCommand{\smuncertaintywithdmat}{O{\beta} m}{\rbk{\triangle #2}_{#1}} 
\NewDocumentCommand{\sphilbfrak}{O{\mathfrak}}{#1} 
\NewDocumentCommand{\sphilb}{O{\mathcal}}{#1} 
\NewDocumentCommand{\splowerhalf}{O{\mathbb{H}}}{#1_{\txtneg}} 
\NewDocumentCommand{\spupperhalf}{O{\mathbb{H}}}{#1_{\txtnonneg}} 
\NewDocumentCommand{\topmetric}{O{d}}{#1} 
\NewDocumentCommand{\vaoutnormal}{O{\widehat}}{#1} 
\newcommand{\category}[1]{\mathop{\mathsf{#1}}} 

\newcommand{\catpresheaf}[1]{\category{PSh}} 
\newcommand{\faadj}[1]{#1^{\ast}} 
\newcommand{\fldcmp}{\fld{C}} 
\newcommand{\fldreal}{\fld{R}} 
\newcommand{\fld}[1]{\mathbb{#1}} 
\newcommand{\lpseq}{\ell} 
\newcommand{\lp}{L} 
\newcommand{\msrcal}[1]{\mathcal{#1}} 

\newcommand{\oaweyl}{\oa{W}} 
\newcommand{\oa}[1]{\mathcal{#1}} 
\newcommand{\opfocksndqntdiff}{d \Gamma} 
\newcommand{\opfockweyl}{W} 
\newcommand{\opspbddlin}[1]{\fun{\mathbb{B}}{#1}} 
\newcommand{\opspecint}[1]{\mathcal{E}} 
\newcommand{\physmean}[1]{\dbk{#1}} 
\newcommand{\ringratint}{\mathbb{Z}} 

\newcommand{\seq}[2]{\if\relax\detokenize{#1}\relax \rbk{#1} \else \rbk{#1}_{#2} \fi} 
\newcommand{\setisomorphism}[1]{\operatorname{Iso}} 
\newcommand{\setone}[1]{\cbk{#1}}
\newcommand{\set}[2]{\left\{#1 \, \middle| \, #2\right\}}
\newcommand{\spfockgamma}{\Gamma} 
\newcommand{\txtasym}{\mathrm{as}} 
\newcommand{\txtbsn}{\mathrm{b}} 
\newcommand{\txtel}{\mathrm{el}} 
\newcommand{\txtfin}{\mathrm{fin}} 
\newcommand{\txtfock}{\mathrm{F}} 
\newcommand{\txtgs}{\mathrm{gs}} 
\newcommand{\txtinteraction}{\mathrm{I}} 
\newcommand{\txtirsingular}{\mathrm{irs}} 
\newcommand{\txtkms}{\mathrm{KMS}} 
\newcommand{\txtmax}{\mathrm{max}} 
\newcommand{\txtneg}{\mathrm{-}} 
\newcommand{\txtnonneg}{\mathrm{+}} 
\newcommand{\txtsym}{\mathrm{s}} 
\newcommand{\txttot}{\mathrm{tot}} 
\newcommand{\txtvac}{\mathrm{vac}} 

\title{Magnetism and infrared divergence in a Hubbard-phonon interacting system}

\author{%
Yoshitsugu Sekine\\{\small\texttt{4429sekine@gmail.com}}%
}

\date{\today}

\begin{document}

\maketitle

\begin{abstract}
We study a finite Hubbard model coupled to a massless phonon field through a form factor common to all sites that satisfies the infrared singular condition. The Hamiltonian has no ground state in the Fock representation. With the unitary transformation of Arai and Hirokawa, the ground states with infrared cutoff define states on the Weyl algebra over the form domain of the inverse dispersion, and they converge to a ground state in a representation inequivalent to the Fock representation. In this representation the ground states are the products of the ground states of the Hubbard model and the phonon vacuum, so that the model has ferromagnetic or unique spin-singlet ground states according to the hopping matrix and the electron number.

\end{abstract}

\setcounter{tocdepth}{2}
\tableofcontents

\section{Introduction}\label{introduction}

Electron-phonon interacting systems are basic models of condensed matter physics, and only a few of their many-electron properties are known rigorously. Freericks and Lieb proved that a general electron-phonon Hamiltonian with finitely many phonon modes and an even number of electrons has a spin-singlet ground state \cite[Theorem 1]{FreericksLieb001}. When the phonons form a quantum field with a massless dispersion relation, the infrared behavior of the coupling decides whether a ground state exists at all. For the generalized spin-boson models of Arai and Hirokawa, which include the model of this paper, the existence and uniqueness of ground states were studied in \cite{AraiHirokawa001,AraiHirokawa002}, and the absence of eigenvectors in massless models in \cite{AraiHirokawaHiroshima001}. For the massless Nelson model without infrared cutoff a ground state exists in a non-Fock representation \cite{AsaoArai019}, and the van Hove model is the exactly solvable prototype of this phenomenon \cite{AsaoArai016,JanDerezinski002}.

In this paper we consider a finite Hubbard model coupled to a massless phonon field. The unitary transformation of Arai and Hirokawa \cite{AraiHirokawa001} is a Lang--Firsov transformation for this model. For mutually orthogonal form factors it changes the on-site coupling from \(U\) to \(U-\rbk{\alpha b_{\kappa}}^2\), an induced attraction, while the hopping terms keep phonon factors (Theorem \ref{thm:hubbard-phonon-lang-firsov}). For a form factor common to all sites the phonons couple only to the total electron number, which is fixed, and the transformation decouples the electrons from the phonons. Our results concern this decoupled case under the infrared singular condition \(\lambda\in\dom\physham[h]_{\txtbsn}^{-1/2}\setminus\dom\physham[h]_{\txtbsn}^{-1}\). The Hamiltonian has no ground state in the Fock representation (Theorem \ref{thm:hubbard-phonon-printed-no-point-spectrum}). The ground states with infrared cutoff converge, as states on the Weyl algebra over \(\dom\physham[h]_{\txtbsn}^{-1/2}\), to a ground state in a representation inequivalent to the Fock representation (Theorems \ref{thm:hubbard-phonon-inequivalent} and \ref{thm:hubbard-phonon-limit-representation}). In this representation the ground states are the products of the ground states of the Hubbard model and the phonon vacuum, so that the theorems of Lieb \cite{ElliottLieb015} and Tasaki \cite{HalTasaki005} give ferromagnetic or unique spin-singlet ground states (Theorem \ref{thm:hubbard-phonon-magnetism}).

The decoupled case is, on the phonon side, a van Hove model, and the removal of the infrared cutoff follows the van Hove model \cite{AsaoArai016}. In particular the phonons do not change the magnetic properties of the electrons, and the coupled case with site-dependent form factors remains open. The finite-temperature counterpart, phonon Bose--Einstein condensation in the same decoupled setting, is treated in \cite{YoshitsuguSekine002}.

In Section \ref{sec:hubbard-phonon-model} we fix the model and derive the Lang--Firsov transformation. In Section \ref{sec:hubbard-phonon-removal} we prove the absence of ground states in the Fock representation and construct the inequivalent representation with a ground state. In Section \ref{sec:hubbard-phonon-ground-states} we apply the theorems on the Hubbard model.

\section{Model and the Lang--Firsov transformation}\label{sec:hubbard-phonon-model}

In this section we define the model with infrared cutoff and transform it by the unitary operator of Arai and Hirokawa.

Let \(\Lambda\) be a finite set, let \(N_{\txtel}\) be an integer with \(1\leq N_{\txtel}\leq2\abs{\Lambda}\), and let \(\physham[h]_{\txtbsn}\) be a nonnegative, injective, and absolutely continuous self-adjoint operator on a separable Hilbert space \(\sphilbfrak{h}_{\txtbsn}\) with \(\inf\opvarspec{\physham[h]_{\txtbsn}}=0\), the dispersion relation of massless phonons; the example in mind is \(\physham[h]_{\txtbsn}=\sqrt{-\laplacian}\) on \(\fun{\lp^{2}}{\fldreal^{d}}\). The Hilbert space is \[\begin{aligned}
\sphilbfrak{H}
=
\sphilbfrak{H}_{\txtel}
\otimes\fun{\spfockgamma_{\txtbsn}}{\sphilbfrak{h}_{\txtbsn}},
\quad
\sphilbfrak{H}_{\txtel}
=
\fun{\lpseq^2}{\Lambda;\fldcmp^2}^{\otimes_{\txtasym}N_{\txtel}},
\end{aligned}\] where \(\fun{\spfockgamma_{\txtbsn}}{\sphilbfrak{h}_{\txtbsn}}=\bigoplus_{n=0}^{\infty}\sphilbfrak{h}_{\txtbsn}^{\otimes_{\txtsym}n}\) is the bosonic Fock space with the vacuum \(\opfockvac[\Psi_{\txtbsn,\txtfock,\txtvac}]\) \cite{AsaoArai016}. For \(f\in\sphilbfrak{h}_{\txtbsn}\), the creation operator \(\faadj{\fun{\opfockan_{\txtfock}}{f}}\) maps an \(n\)-particle vector \(\Psi^{\rbk{n}}\) to \(\sqrt{n+1}\) times the symmetrization of \(f\otimes\Psi^{\rbk{n}}\), the annihilation operator \(\fun{\opfockan_{\txtfock}}{f}\) is its adjoint, and Segal's field operator \(\fun{\opfocksegal_{\txtfock}}{f}\) is the closure of \(2^{-1/2}\rbk{\fun{\opfockan_{\txtfock}}{f}+\faadj{\fun{\opfockan_{\txtfock}}{f}}}\) on the finite-particle vectors, which is self-adjoint; the map \(f\mapsto\fun{\opfocksegal_{\txtfock}}{f}\) is real linear. For a self-adjoint operator \(T\) on \(\sphilbfrak{h}_{\txtbsn}\), \(\fun{\opfocksndqntdiff_{\txtbsn}}{T}\) is its second quantization, and \(\opfocknumber_{\txtbsn}=\fun{\opfocksndqntdiff_{\txtbsn}}{\idone}\) is the number operator \cite{AsaoArai016}. The Hamiltonian is \begin{equation}\label{eq:hubbard-phonon-hamiltonian}
\begin{aligned}
H
&=
H_{\txtel}\otimes\idone+\idone\otimes H_{\txtbsn}
+\alpha\sum_{x \in \Lambda} n_x \otimes \fun{\opfocksegal_{\txtfock}}{\lambda_x},
\\ 
H_{\txtel}
&=
\sum_{x,y\in\Lambda}\sum_{s=\pm}t_{x,y}\faadj{c_{x,s}}c_{y,s} + U
\sum_{x \in \Lambda} n_{x, +} n_{x, -},
\quad
H_{\txtbsn}
=
\fun{\opfocksndqntdiff_{\txtbsn}}{\physham[h]_{\txtbsn}},
\end{aligned}
\end{equation} where \(c_{x,s}\) are the electron annihilation operators restricted to \(\sphilbfrak{H}_{\txtel}\), \(n_{x,s}=\faadj{c_{x,s}}c_{x,s}\), \(n_x=n_{x,+}+n_{x,-}\), \(\rbk{t_{x,y}}_{x,y\in\Lambda}\) is a self-adjoint hopping matrix, \(U>0\), \(\alpha\in\fldreal\), \(\opfocksegal_{\txtfock}\) is Segal's field operator, and \(\lambda_x\in\sphilbfrak{h}_{\txtbsn}\). On \(\sphilbfrak{H}_{\txtel}\), \(\sum_{x\in\Lambda}n_x=N_{\txtel}\). We write \(\fun{E_0}{A}=\inf\opvarspec{A}\) for a self-adjoint operator \(A\).

\begin{defn}[Assumption]\label{def:hubbard-phonon-printed-infrared-singular}
The vectors $\lambda_x$, $x\in\Lambda$, belong to $\dom\physham[h]_{\txtbsn}^{-1/2}$,
no partial sum of them belongs to $\dom\physham[h]_{\txtbsn}^{-1}$, and
$\bkt{\lambda_x}{\lambda_y}_{\sphilbfrak{h}_{\txtbsn}}$ and
$\bkt{\physham[h]_{\txtbsn}^{-1/2}\lambda_x}{\physham[h]_{\txtbsn}^{-1/2}\lambda_y}_{\sphilbfrak{h}_{\txtbsn}}$
are real for all $x,y\in\Lambda$.
\end{defn}

The condition \(\lambda_x\notin\dom\physham[h]_{\txtbsn}^{-1}\) is the infrared singular condition. Since, for \(f\in\dom\physham[h]_{\txtbsn}^{-1/2}\), \begin{equation}\label{eq:hubbard-phonon-field-bound}
\begin{aligned}
\norm{\fun{\opfocksegal_{\txtfock}}{f}\Psi}
\leq
2^{-1/2}\rbk{2\norm{\physham[h]_{\txtbsn}^{-1/2}f}_{\sphilbfrak{h}_{\txtbsn}}\norm{H_{\txtbsn}^{1/2}\Psi}+\norm{f}_{\sphilbfrak{h}_{\txtbsn}}\norm{\Psi}},
\end{aligned}
\end{equation} the interaction is infinitesimally bounded with respect to \(H_{\txtel}\otimes\idone+\idone\otimes H_{\txtbsn}\). Hence, by the Kato--Rellich theorem, \(H\) is self-adjoint on \(\dom\rbk{\idone\otimes H_{\txtbsn}}\), bounded from below, and essentially self-adjoint on \(\sphilbfrak{D}_{\physham}=\sphilbfrak{H}_{\txtel}\algotimes\fun{\spfockgamma_{\txtbsn,\txtfin}}{\dom\physham[h]_{\txtbsn}}\), where \(\fun{\spfockgamma_{\txtbsn,\txtfin}}{\dom\physham[h]_{\txtbsn}}\) is the linear span of the finite-particle vectors with entries in \(\dom\physham[h]_{\txtbsn}\) \cite[Proposition 1.1, pp.~460--461]{AraiHirokawa001}.

\begin{defn}[Assumption]\label{def:hubbard-phonon-printed-regularization}
For $\kappa>0$, $\lambda_{x,\kappa}\in\dom\physham[h]_{\txtbsn}^{-1/2}\cap\dom\physham[h]_{\txtbsn}^{-1}$,
the inner products of Assumption \ref{def:hubbard-phonon-printed-infrared-singular}
and $\bkt{\physham[h]_{\txtbsn}^{-1}\lambda_{x,\kappa}}{\physham[h]_{\txtbsn}^{-1}\lambda_{y,\kappa}}_{\sphilbfrak{h}_{\txtbsn}}$
are real, and $\lambda_{x,\kappa}\to\lambda_x$ and
$\physham[h]_{\txtbsn}^{-1/2}\lambda_{x,\kappa}\to\physham[h]_{\txtbsn}^{-1/2}\lambda_x$
as $\kappa\to0$.
\end{defn}

The Hamiltonian \(H_{\kappa}\) is \eqref{eq:hubbard-phonon-hamiltonian} with \(\lambda_{x,\kappa}\) in place of \(\lambda_x\), and \(H_{\txtinteraction,\kappa}=\sum_xn_x\otimes\fun{\opfocksegal_{\txtfock}}{\lambda_{x,\kappa}}\). The field estimate \eqref{eq:hubbard-phonon-field-bound} gives \(\norm{\rbk{H_{\kappa}-H}\Psi}\to0\) for \(\Psi\in\sphilbfrak{D}_{\physham}\). Since \(\sphilbfrak{D}_{\physham}\) is a common core and \(\rbk{H_{\kappa}-\zeta}^{-1}-\rbk{H-\zeta}^{-1}=\rbk{H_{\kappa}-\zeta}^{-1}\rbk{H-H_{\kappa}}\rbk{H-\zeta}^{-1}\) with \(\norm{\rbk{H_{\kappa}-\zeta}^{-1}}\leq\abs{\opimag\zeta}^{-1}\), it follows that \(H_{\kappa}\to H\) in the strong resolvent sense. Following \cite{AraiHirokawa001}, put \(S_{\kappa}=\sum_{x\in\Lambda}n_x\otimes\fun{\opfocksegal_{\txtfock}}{\imunit\physham[h]_{\txtbsn}^{-1}\lambda_{x,\kappa}}\), which is self-adjoint by Nelson's analytic vector theorem, and \(V_{\kappa}=\napiernum^{\imunit\alpha S_{\kappa}}\).

\begin{prop}[{\cite[Lemma 5.1, p.~495]{AraiHirokawa001}}]\label{prop:hubbard-phonon-printed-transform}
For $\kappa>0$,
$$\begin{aligned}
V_{\kappa}\rbk{\idone\otimes H_{\txtbsn}}V_{\kappa}^{-1}
=
\idone\otimes H_{\txtbsn}+\alpha H_{\txtinteraction,\kappa}+\alpha^2R_{\kappa}\otimes\idone,
\quad
R_{\kappa}
=
\frac{1}{2}\sum_{x,y\in\Lambda}
\bkt{\physham[h]_{\txtbsn}^{-1/2}\lambda_{x,\kappa}}{\physham[h]_{\txtbsn}^{-1/2}\lambda_{y,\kappa}}_{\sphilbfrak{h}_{\txtbsn}}n_xn_y.
\end{aligned}$$
\end{prop}

\begin{proof}
The vectors of $\sphilbfrak{D}_{\physham}$ are analytic for $S_{\kappa}$, and on
them the commutators of $S_{\kappa}$ with $\idone\otimes H_{\txtbsn}$ are
$-\imunit H_{\txtinteraction,\kappa}$, then $-2R_{\kappa}\otimes\idone$, and then
$0$, by the canonical commutation relations and the reality in
Assumption \ref{def:hubbard-phonon-printed-regularization}.
The commutator expansion of $V_{\kappa}\rbk{\idone\otimes H_{\txtbsn}}V_{\kappa}^{-1}$
therefore terminates on this core.
\end{proof}

\begin{thm}[Lang--Firsov transformation]\label{thm:hubbard-phonon-lang-firsov}
For $\kappa>0$,
\begin{equation}\label{eq:hubbard-phonon-lang-firsov}
\begin{aligned}
V_{\kappa}^{-1}H_{\kappa}V_{\kappa}
&=
\sum_{x,y\in\Lambda}\sum_{s=\pm}t_{x,y}\faadj{c_{x,s}}c_{y,s}\otimes
\napiernum^{-\imunit\alpha\fun{\opfocksegal_{\txtfock}}{\imunit\physham[h]_{\txtbsn}^{-1}
\lambda_{x,\kappa}}}
\napiernum^{\imunit\alpha\fun{\opfocksegal_{\txtfock}}{\imunit\physham[h]_{\txtbsn}^{-1}
\lambda_{y,\kappa}}}
\\ 
&\quad
+\rbk{U\sum_{x\in\Lambda}n_{x,+}n_{x,-}-\alpha^2R_{\kappa}}\otimes\idone
+\idone\otimes H_{\txtbsn}.
\end{aligned}
\end{equation}
If, moreover,
\begin{equation}\label{eq:hubbard-phonon-orthogonal-form-factors}
\begin{aligned}
\bkt{\physham[h]_{\txtbsn}^{-1/2} \lambda_{x,\kappa}}
{\physham[h]_{\txtbsn}^{-1/2} \lambda_{y,\kappa}}_{\sphilbfrak{h}_{\txtbsn}}
=
\kroneckerdelta_{x,y} b_{\kappa}^2
\quad
\rbk{x,y\in\Lambda}
\end{aligned}
\end{equation}
with $b_{\kappa}\geq0$, then
$U\sum_xn_{x,+}n_{x,-}-\alpha^2R_{\kappa}=\rbk{U-\rbk{\alpha b_{\kappa}}^2}\sum_xn_{x,+}n_{x,-}-\frac{\rbk{\alpha b_{\kappa}}^2}{2}N_{\txtel}$.
\end{thm}

\begin{proof}
Let $P_m$, $m\in\setone{0,1,2}^{\Lambda}$, be the joint spectral projections of
the commuting operators $n_z$, and put $P_m=0$ for
$m\in\ringratint^{\Lambda}\setminus\setone{0,1,2}^{\Lambda}$.
On the range of $P_m\otimes\idone$, $S_{\kappa}$ acts as
$\idone\otimes\fun{\opfocksegal_{\txtfock}}{\imunit\sum_zm_z\physham[h]_{\txtbsn}^{-1}\lambda_{z,\kappa}}$,
so that
$V_{\kappa}\rbk{P_m\otimes\idone}=P_m\otimes\napiernum^{\imunit\alpha\fun{\opfocksegal_{\txtfock}}{\imunit\sum_zm_z\physham[h]_{\txtbsn}^{-1}\lambda_{z,\kappa}}}$.
Since $\faadj{c_{x,s}}c_{y,s}P_m=P_{m+\delta_x-\delta_y}\faadj{c_{x,s}}c_{y,s}P_m$,
where $\delta_z$ is the indicator function of $\setone{z}$, and since the imaginary
parts of the inner products of the vectors
$\imunit\physham[h]_{\txtbsn}^{-1}\lambda_{z,\kappa}$ vanish by
Assumption \ref{def:hubbard-phonon-printed-regularization}, the Weyl relations
\cite[Proposition 5.2.4(2), p.~13]{BratteliRobinson004} give the hopping terms
of \eqref{eq:hubbard-phonon-lang-firsov} after summation over $m$.
The operators $n_{x,+}n_{x,-}$ and $R_{\kappa}$ commute with every $P_m$, and
Proposition \ref{prop:hubbard-phonon-printed-transform} gives the remaining terms.
Under \eqref{eq:hubbard-phonon-orthogonal-form-factors},
$R_{\kappa}=\frac{b_{\kappa}^2}{2}\sum_xn_x^2$ and
$n_x^2=n_x+2n_{x,+}n_{x,-}$.
\end{proof}

Under \eqref{eq:hubbard-phonon-orthogonal-form-factors} the on-site coupling becomes \(U-\rbk{\alpha b_{\kappa}}^2\), which is negative for large \(\abs{\alpha}\), while the phonon factors of the off-diagonal hopping terms in \eqref{eq:hubbard-phonon-lang-firsov} do not cancel when \(\alpha b_{\kappa}\neq0\). The rest of the paper treats the case in which they cancel.

\section{Removal of the infrared cutoff}\label{sec:hubbard-phonon-removal}

In this section we prove the absence of ground states in the Fock representation and construct an inequivalent representation in which the limit of the cutoff ground states is a ground state; the argument follows the van Hove model \cite{AsaoArai016,JanDerezinski002}.

\begin{defn}[Assumption of a common form factor]\label{def:hubbard-phonon-common-form-factor}
There are vectors $\lambda$ and $\lambda_{\kappa}$, $\kappa>0$, such that
\begin{equation}\label{eq:hubbard-phonon-common-form-factor}
\begin{aligned}
\lambda_x=\lambda,
\quad
\lambda_{x,\kappa}=\lambda_{\kappa}
\end{aligned}
\end{equation}
for all $x\in\Lambda$ and $\kappa>0$.
\end{defn}

If \(\abs{\Lambda}\geq2\), \eqref{eq:hubbard-phonon-common-form-factor} and \eqref{eq:hubbard-phonon-orthogonal-form-factors} with \(b_{\kappa}>0\) exclude each other. From here on we assume \(\alpha\neq0\) and \eqref{eq:hubbard-phonon-common-form-factor}. Since \(\sum_xn_x=N_{\txtel}\), Theorem \ref{thm:hubbard-phonon-lang-firsov} gives \begin{equation}\label{eq:hubbard-phonon-decoupled-unitary}
\begin{aligned}
V_{\kappa}=\idone\otimes\napiernum^{\imunit\fun{\opfocksegal_{\txtfock}}{\imunit\alpha N_{\txtel}\physham[h]_{\txtbsn}^{-1}\lambda_{\kappa}}}
\end{aligned}
\end{equation} and \begin{equation}\label{eq:hubbard-phonon-common-decoupled}
\begin{aligned}
H_{\kappa}
=
V_{\kappa}\hat{H}_{\kappa}V_{\kappa}^{-1},
\quad
\hat{H}_{\kappa}
=
\hat{H}_{\txtel,\kappa}\otimes\idone+\idone\otimes H_{\txtbsn},
\quad
\hat{H}_{\txtel,\kappa}
=
H_{\txtel}-\frac{\alpha^2N_{\txtel}^2}{2}
\norm{\physham[h]_{\txtbsn}^{-1/2}\lambda_{\kappa}}_{\sphilbfrak{h}_{\txtbsn}}^2.
\end{aligned}
\end{equation} We define \(\hat{H}_{\txtel,0}\) and \(\hat{H}_{0}\) by \eqref{eq:hubbard-phonon-common-decoupled} with \(\lambda\) in place of \(\lambda_{\kappa}\). Since \(\physham[h]_{\txtbsn}\) has no eigenvalues, the ground states of \(H_{\kappa}\) are the vectors \(\Psi_{\txtgs,\kappa}=V_{\kappa}\rbk{\Psi_{\txtel}\otimes\opfockvac[\Psi_{\txtbsn,\txtfock,\txtvac}]}\) with ground states \(\Psi_{\txtel}\) of \(H_{\txtel}\). Since \(\napiernum^{\imunit\fun{\opfocksegal_{\txtfock}}{\imunit g}}\opfocknumber_{\txtbsn}\napiernum^{-\imunit\fun{\opfocksegal_{\txtfock}}{\imunit g}}=\opfocknumber_{\txtbsn}+\fun{\opfocksegal_{\txtfock}}{g}+\frac{1}{2}\norm{g}^2\) for \(g\in\sphilbfrak{h}_{\txtbsn}\) \cite{AsaoArai016}, and the field operator has vanishing vacuum expectation, \eqref{eq:hubbard-phonon-decoupled-unitary} gives \[\begin{aligned}
\bkt{\Psi_{\txtgs,\kappa}}{\rbk{\idone\otimes\opfocknumber_{\txtbsn}}\Psi_{\txtgs,\kappa}}
=
\frac{\alpha^2N_{\txtel}^2}{2}
\norm{\physham[h]_{\txtbsn}^{-1}\lambda_{\kappa}}_{\sphilbfrak{h}_{\txtbsn}}^2
\end{aligned}\] for normalized \(\Psi_{\txtel}\), and this diverges as \(\kappa\to0\) by Assumption \ref{def:hubbard-phonon-printed-infrared-singular}.

\subsection{Absence of ground states in the Fock representation}\label{absence-of-ground-states-in-the-fock-representation}

In this subsection we show that \(H\) has no eigenvector and that the cutoff ground states converge weakly to \(0\); on the phonon side these are properties of the van Hove model \cite{AsaoArai016}.

By \eqref{eq:hubbard-phonon-common-form-factor} and \(\sum_xn_x=N_{\txtel}\), \begin{equation}\label{eq:hubbard-phonon-van-hove-form}
\begin{aligned}
H
=
H_{\txtel}\otimes\idone
+\idone\otimes\rbk{H_{\txtbsn}+\fun{\opfocksegal_{\txtfock}}{\alpha N_{\txtel}\lambda}},
\end{aligned}
\end{equation} and the second operator is the van Hove Hamiltonian with the form factor \(\alpha N_{\txtel}\lambda\in\dom\physham[h]_{\txtbsn}^{-1/2}\setminus\dom\physham[h]_{\txtbsn}^{-1}\).

\begin{thm}\label{thm:hubbard-phonon-printed-no-point-spectrum}
The operator $H$ has no eigenvalues; in particular it has no ground state.
\end{thm}

\begin{proof}
The van Hove Hamiltonian in \eqref{eq:hubbard-phonon-van-hove-form} has no
eigenvalues, since $\physham[h]_{\txtbsn}$ is absolutely continuous and
$\alpha N_{\txtel}\lambda\notin\dom\physham[h]_{\txtbsn}^{-1}$ \cite{AsaoArai016}.
An orthonormal eigenbasis of the finite-dimensional operator $H_{\txtel}$ with
eigenvalues $E_1,\ldots,E_{\dim\sphilbfrak{H}_{\txtel}}$ decomposes $H$ into the
direct sum of the operators
$E_j+H_{\txtbsn}+\fun{\opfocksegal_{\txtfock}}{\alpha N_{\txtel}\lambda}$.
Each summand has no eigenvalues, and the eigenvalues of a finite direct sum are
those of its summands, so that $H$ has no eigenvalues.
\end{proof}

The cutoff ground states have no limit in \(\sphilbfrak{H}\) either. By \eqref{eq:hubbard-phonon-decoupled-unitary}, \[\begin{aligned}
\Psi_{\txtgs,\kappa}
=
\Psi_{\txtel}\otimes\napiernum^{\imunit\fun{\opfocksegal_{\txtfock}}{\imunit\alpha N_{\txtel}\physham[h]_{\txtbsn}^{-1}\lambda_{\kappa}}}\opfockvac[\Psi_{\txtbsn,\txtfock,\txtvac}],
\end{aligned}\] and the second factor is the ground state of the van Hove Hamiltonian with the form factor \(\alpha N_{\txtel}\lambda_{\kappa}\). Since \(\norm{\physham[h]_{\txtbsn}^{-1}\lambda_{\kappa}}\to\infty\), it converges weakly to \(0\) as \(\kappa\to0\) \cite{AsaoArai016}, and hence \(\Psi_{\txtgs,\kappa}\to0\) weakly.

\subsection{The inequivalent representation and its ground state}\label{the-inequivalent-representation-and-its-ground-state}

In this subsection we describe the limit of the cutoff ground states on the Weyl algebra over \(\dom\physham[h]_{\txtbsn}^{-1/2}\) and show that the limit representation is inequivalent to the Fock representation.

For a complex linear subspace \(\sphilbfrak{D}\) of \(\sphilbfrak{h}_{\txtbsn}\), the form \(\opimag\bkt{f}{g}_{\sphilbfrak{h}_{\txtbsn}}\) is a nondegenerate symplectic form on \(\sphilbfrak{D}\), and we let \(\fun{\oaweyl}{\sphilbfrak{D}}\) be the Weyl algebra over it with generators \(\fun{\opfockweyl}{f}\) \cite[Theorem 5.2.8, p.~19]{BratteliRobinson004}. With \(\oa{A}_{\txtel}=\opspbddlin{\sphilbfrak{H}_{\txtel}}\) put \begin{equation}\label{eq:hubbard-phonon-limit-algebras}
\begin{aligned}
\oa{A}_{\mathrm{irr}}
=
\oa{A}_{\txtel}\otimes\fun{\oaweyl}{\sphilbfrak{h}_{\txtbsn}},
\quad
\oa{A}_{\txtirsingular}
=
\oa{A}_{\txtel}\otimes\fun{\oaweyl}{\dom\physham[h]_{\txtbsn}^{-1/2}}.
\end{aligned}
\end{equation} By \cite[Theorem 5.2.8, p.~19]{BratteliRobinson004} the canonical map \(\fun{\oaweyl}{\dom\physham[h]_{\txtbsn}^{-1/2}}\to\fun{\oaweyl}{\sphilbfrak{h}_{\txtbsn}}\) is an injective \(\ast\)-homomorphism, and we regard \(\oa{A}_{\txtirsingular}\) as a subalgebra of \(\oa{A}_{\mathrm{irr}}\). The same theorem gives representations \(\pi_{\mathrm{irr}}\) of \(\oa{A}_{\mathrm{irr}}\) and \(\pi_{\txtirsingular}\) of \(\oa{A}_{\txtirsingular}\) on \(\sphilbfrak{H}\) with \begin{equation}\label{eq:hubbard-phonon-limit-representations}
\begin{aligned}
\fun{\pi_{\mathrm{irr}}}{A\otimes\fun{\opfockweyl}{f}}
&=
A\otimes\napiernum^{\imunit\fun{\opfocksegal_{\txtfock}}{f}},
\\ 
\fun{\pi_{\txtirsingular}}{A\otimes\fun{\opfockweyl}{f}}
&=
\napiernum^{-\imunit\alpha N_{\txtel}\opreal\bkt{\physham[h]_{\txtbsn}^{-1/2}\lambda}{\physham[h]_{\txtbsn}^{-1/2}f}_{\sphilbfrak{h}_{\txtbsn}}}
A\otimes\napiernum^{\imunit\fun{\opfocksegal_{\txtfock}}{f}},
\end{aligned}
\end{equation} since the phase is real linear in \(f\); \(\pi_{\mathrm{irr}}\) is the Fock representation. For \(\kappa>0\) put \(\fun{\pi_{\kappa}}{X}=V_{\kappa}^{-1}\fun{\pi_{\mathrm{irr}}}{X}V_{\kappa}\), \(\fun{\alpha_{\kappa,t}}{A}=\napiernum^{\imunit tH_{\kappa}}A\napiernum^{-\imunit tH_{\kappa}}\), \(\fun{\alpha_{0,t}}{A}=\napiernum^{\imunit t\hat{H}_{0}}A\napiernum^{-\imunit t\hat{H}_{0}}\), and, for a normalized ground state \(\Psi_{\txtgs,\kappa}\) of \(H_{\kappa}\), \(\fun{\oastate[\psi_{\txtgs,\kappa}]}{X}=\bkt{\Psi_{\txtgs,\kappa}}{\fun{\pi_{\mathrm{irr}}}{X}\Psi_{\txtgs,\kappa}}\).

\begin{prop}\label{prop:hubbard-phonon-limit-map}
For $A\in\oa{A}_{\txtel}$, $f\in\sphilbfrak{h}_{\txtbsn}$, and $\kappa>0$,
\begin{equation}\label{eq:hubbard-phonon-pi-kappa-weyl}
\begin{aligned}
\fun{\pi_{\kappa}}{A\otimes\fun{\opfockweyl}{f}}
=
\napiernum^{-\imunit\alpha N_{\txtel}\opreal\bkt{f}{\physham[h]_{\txtbsn}^{-1}\lambda_{\kappa}}_{\sphilbfrak{h}_{\txtbsn}}}
A\otimes\napiernum^{\imunit\fun{\opfocksegal_{\txtfock}}{f}},
\end{aligned}
\end{equation}
and $\fun{\pi_{\kappa}}{X}\to\fun{\pi_{\txtirsingular}}{X}$ in norm as $\kappa\to0$
for every $X\in\oa{A}_{\txtirsingular}$.
\end{prop}

\begin{proof}
The formula follows from \eqref{eq:hubbard-phonon-decoupled-unitary} and the Weyl
relations.
For $f\in\dom\physham[h]_{\txtbsn}^{-1/2}$ the phase equals
$\opreal\bkt{\physham[h]_{\txtbsn}^{-1/2}f}{\physham[h]_{\txtbsn}^{-1/2}\lambda_{\kappa}}$
and converges to the phase of \eqref{eq:hubbard-phonon-limit-representations} by
Assumption \ref{def:hubbard-phonon-printed-regularization}.
The linear span of the generators is dense in $\oa{A}_{\txtirsingular}$, and
every representation $\pi$ satisfies $\norm{\fun{\pi}{X}}\leq\norm{X}$.
\end{proof}

Both \(\alpha_{\kappa,t}\) and \(\alpha_{0,t}\) map the generators to scalar multiples of generators, by \eqref{eq:hubbard-phonon-common-decoupled} and \eqref{eq:hubbard-phonon-pi-kappa-weyl}, and \(\napiernum^{\imunit t\physham[h]_{\txtbsn}}\) preserves \(\dom\physham[h]_{\txtbsn}^{-1/2}\). Hence \(\alpha_{\kappa,t}\) preserves \(\fun{\pi_{\mathrm{irr}}}{\oa{A}_{\mathrm{irr}}}\), \(\alpha_{0,t}\) preserves \(\fun{\pi_{\txtirsingular}}{\oa{A}_{\txtirsingular}}\), and \(\oastate[\psi_{\txtgs,\kappa}]\) is a ground state for \(\alpha_{\kappa}\) in the sense that \(H_{\kappa}-\fun{E_0}{H_{\kappa}}\geq0\) implements \(\alpha_{\kappa}\) and annihilates \(\Psi_{\txtgs,\kappa}\).

\begin{thm}\label{thm:hubbard-phonon-inequivalent}
The representation $\pi_{\txtirsingular}$ is not unitarily equivalent to the
restriction of the Fock representation $\pi_{\mathrm{irr}}$ to
$\oa{A}_{\txtirsingular}$.
\end{thm}

\begin{proof}
We first compare irreducible representations of the phonon part and then
pass to the direct sums.
On $\idone\otimes\fun{\oaweyl}{\dom\physham[h]_{\txtbsn}^{-1/2}}$ the two
representations are direct sums of $\dim\sphilbfrak{H}_{\txtel}$ copies of the
irreducible representations $\rho_{\txtirsingular}$ and $\rho_{\mathrm{irr}}$ on
$\fun{\spfockgamma_{\txtbsn}}{\sphilbfrak{h}_{\txtbsn}}$ given by
$$\begin{aligned}
\fun{\rho_{\txtirsingular}}{\fun{\opfockweyl}{f}}
&=
\napiernum^{-\imunit\alpha N_{\txtel}\opreal\bkt{\physham[h]_{\txtbsn}^{-1/2}\lambda}{\physham[h]_{\txtbsn}^{-1/2}f}_{\sphilbfrak{h}_{\txtbsn}}}
\napiernum^{\imunit\fun{\opfocksegal_{\txtfock}}{f}},\\
\fun{\rho_{\mathrm{irr}}}{\fun{\opfockweyl}{f}}
&=
\napiernum^{\imunit\fun{\opfocksegal_{\txtfock}}{f}}.
\end{aligned}$$
By Stone's theorem a unitary operator intertwines $\rho_{\txtirsingular}$ and
$\rho_{\mathrm{irr}}$ exactly when it intertwines the generators
$\fun{\opfocksegal_{\txtfock}}{f}-\alpha N_{\txtel}\opreal\bkt{\physham[h]_{\txtbsn}^{-1/2}\lambda}{\physham[h]_{\txtbsn}^{-1/2}f}_{\sphilbfrak{h}_{\txtbsn}}$
and $\fun{\opfocksegal_{\txtfock}}{f}$, that is, by taking $f$ and $\imunit f$,
the Bogoliubov translation
$\fun{\opfockan_{\txtfock}}{f}-2^{-1/2}\alpha N_{\txtel}\bkt{\physham[h]_{\txtbsn}^{-1/2}f}{\physham[h]_{\txtbsn}^{-1/2}\lambda}_{\sphilbfrak{h}_{\txtbsn}}$
of the annihilation operators and $\fun{\opfockan_{\txtfock}}{f}$.
Since $\alpha N_{\txtel}\lambda\notin\dom\physham[h]_{\txtbsn}^{-1}$ by
Assumption \ref{def:hubbard-phonon-printed-infrared-singular}, this translation
is not equivalent to the Fock representation of the canonical commutation
relations \cite{AsaoArai016}, and hence no unitary operator intertwines
$\rho_{\txtirsingular}$ and $\rho_{\mathrm{irr}}$.

We next show that every bounded operator $U$ intertwining the two direct sums
vanishes.
For unit vectors $e_i,e_j\in\sphilbfrak{H}_{\txtel}$ define the bounded
operator $T$ on $\fun{\spfockgamma_{\txtbsn}}{\sphilbfrak{h}_{\txtbsn}}$ by
$$\begin{aligned}
\bkt{\eta}{T\xi}
=
\bkt{e_i\otimes\eta}{U\rbk{e_j\otimes\xi}},
\quad
\eta,\xi\in\fun{\spfockgamma_{\txtbsn}}{\sphilbfrak{h}_{\txtbsn}}.
\end{aligned}$$
Then $T\fun{\rho_{\txtirsingular}}{X}=\fun{\rho_{\mathrm{irr}}}{X}T$ for every
$X\in\fun{\oaweyl}{\dom\physham[h]_{\txtbsn}^{-1/2}}$.
Then $\faadj{T}T$ commutes with $\rho_{\txtirsingular}$ and $T\faadj{T}$ with
$\rho_{\mathrm{irr}}$, so that both equal the same constant $c\geq0$ by
irreducibility.
For $c>0$ the operator $T$ is $c^{1/2}$ times a unitary operator intertwining
$\rho_{\txtirsingular}$ and $\rho_{\mathrm{irr}}$, and the first step excludes
this case; hence $c=0$ and $T=0$.
All matrix entries of $U$ vanish, so that $U=0$ and in particular no unitary
operator intertwines $\pi_{\txtirsingular}$ and the restriction of
$\pi_{\mathrm{irr}}$.
\end{proof}

Fix a normalized ground state \(\Psi_{\txtel}\) of \(H_{\txtel}\), let \(\Psi_{\txtgs,\kappa}=V_{\kappa}\rbk{\Psi_{\txtel}\otimes\opfockvac[\Psi_{\txtbsn,\txtfock,\txtvac}]}\), and put \(\Psi_{\txtgs,0}=\Psi_{\txtel}\otimes\opfockvac[\Psi_{\txtbsn,\txtfock,\txtvac}]\) and \(\fun{\oastate[\psi_{\txtgs,0}]}{X}=\bkt{\Psi_{\txtgs,0}}{\fun{\pi_{\txtirsingular}}{X}\Psi_{\txtgs,0}}\) for \(X\in\oa{A}_{\txtirsingular}\).

\begin{thm}\label{thm:hubbard-phonon-limit-representation}
For $X\in\oa{A}_{\txtirsingular}$ and $t\in\fldreal$,
$$\begin{aligned}
\lim_{\kappa\to0}\fun{\oastate[\psi_{\txtgs,\kappa}]}{X}
=
\fun{\oastate[\psi_{\txtgs,0}]}{X},
\quad
\lim_{\kappa\to0}V_{\kappa}^{-1}\fun{\alpha_{\kappa,t}}{\fun{\pi_{\mathrm{irr}}}{X}}V_{\kappa}
=
\fun{\alpha_{0,t}}{\fun{\pi_{\txtirsingular}}{X}}
\end{aligned}$$
in norm.
The operator
$\hat{H}_{0}-\fun{E_0}{\hat{H}_{0}}=\rbk{H_{\txtel}-\fun{E_0}{H_{\txtel}}}\otimes\idone+\idone\otimes H_{\txtbsn}$
is nonnegative, implements $\alpha_{0}$, and annihilates $\Psi_{\txtgs,0}$.
In particular $\oastate[\psi_{\txtgs,0}]$ is a ground state for $\alpha_{0}$ in
the representation $\pi_{\txtirsingular}$, which is inequivalent to the Fock
representation.
\end{thm}

\begin{proof}
Since $\fun{\oastate[\psi_{\txtgs,\kappa}]}{X}=\bkt{\Psi_{\txtgs,0}}{\fun{\pi_{\kappa}}{X}\Psi_{\txtgs,0}}$
and
$V_{\kappa}^{-1}\fun{\alpha_{\kappa,t}}{\fun{\pi_{\mathrm{irr}}}{X}}V_{\kappa}=\napiernum^{\imunit t\hat{H}_{0}}\fun{\pi_{\kappa}}{X}\napiernum^{-\imunit t\hat{H}_{0}}$
by \eqref{eq:hubbard-phonon-common-decoupled}, as $\hat{H}_{\kappa}-\hat{H}_{0}$ is
a scalar, both limits follow from Proposition \ref{prop:hubbard-phonon-limit-map}.
The last statement follows from $\fun{E_0}{H_{\txtbsn}}=0$.
\end{proof}

\section{Magnetism in the inequivalent representation}\label{sec:hubbard-phonon-ground-states}

In this section we combine Theorem \ref{thm:hubbard-phonon-limit-representation} with two theorems on the Hubbard model.

Assume that \(\rbk{t_{x,y}}_{x,y\in\Lambda}\) is real symmetric and that \(\Lambda\) is connected by the bonds \(\setone{x,y}\) with \(t_{x,y}\neq0\). Let \(\hat{\mathbb{S}}_{\txttot}\) be the total spin operator of the electrons, \(S_{\txttot}\rbk{S_{\txttot}+1}\) the eigenvalues of \(\hat{\mathbb{S}}_{\txttot}^2\), and \(S_{\txtmax}=\min\setone{N_{\txtel},2\abs{\Lambda}-N_{\txtel}}/2\) the maximal total spin. The Hubbard model is ferromagnetic if all its ground states have \(S_{\txttot}=S_{\txtmax}\).

\begin{thm}[{\cite{ElliottLieb015}, \cite[Theorem 3.5]{HalTasaki005}}]\label{thm:hubbard-phonon-printed-lieb}
Let $\Lambda=A\cup B$ with $A\cap B=\emptyset$ and $\abs{A}=\abs{B}$, let
$t_{x,y}=0$ if $x,y\in A$ or $x,y\in B$, and let $N_{\txtel}=\abs{\Lambda}$.
Then the ground state of $H_{\txtel}$ is unique and has $S_{\txttot}=0$.
\end{thm}

\begin{thm}[{\cite[Theorem 5.1]{HalTasaki005}}]\label{thm:hubbard-phonon-printed-tasaki}
Let $t_{x,y}=t_0t_xt_y$ with $t_0>0$ and $t_x>0$, and let
$N_{\txtel}=\abs{\Lambda}-1$.
Then $H_{\txtel}$ is ferromagnetic, and its ground states are nondegenerate apart
from the $\rbk{2S_{\txtmax}+1}$-fold spin degeneracy.
\end{thm}

\begin{thm}\label{thm:hubbard-phonon-magnetism}
Assume \eqref{eq:hubbard-phonon-common-form-factor}, $\alpha\neq0$, and
Assumptions \ref{def:hubbard-phonon-printed-infrared-singular} and
\ref{def:hubbard-phonon-printed-regularization}.
The eigenspace of $\hat{H}_{0}$ for the eigenvalue $\fun{E_0}{\hat{H}_{0}}$ is
$\set{\Psi_{\txtel}\otimes\opfockvac[\Psi_{\txtbsn,\txtfock,\txtvac}]}{\Psi_{\txtel}\in\Ker\rbk{H_{\txtel}-\fun{E_0}{H_{\txtel}}}}$,
and the electron observables act on the first factor.
Hence, in the representation $\pi_{\txtirsingular}$, the model has a unique
spin-singlet ground state under the hypotheses of
Theorem \ref{thm:hubbard-phonon-printed-lieb} and ferromagnetic ground states
under those of Theorem \ref{thm:hubbard-phonon-printed-tasaki}, for every
$\alpha\neq0$.
\end{thm}

\begin{proof}
Since $\physham[h]_{\txtbsn}$ has no eigenvalues, $\opfockvac[\Psi_{\txtbsn,\txtfock,\txtvac}]$
spans the kernel of $H_{\txtbsn}$, and $\hat{H}_{\txtel,0}-H_{\txtel}$ is a scalar.
Theorems \ref{thm:hubbard-phonon-printed-lieb} and
\ref{thm:hubbard-phonon-printed-tasaki} then apply to the first factor.
\end{proof}

\section{AI statement}\label{ai-statement}

Large language models were used to translate and reorganize the author's draft, to proofread the English text, to check cross-references and bibliographic citations, and to check the details of the proofs for errors and inconsistencies. The author reviewed and verified all suggested changes and remains fully responsible for the mathematical content and the text of this paper.

\bibliography{myref.bib}

\end{document}